\documentclass[a4paper,12pt]{article}
\usepackage{jheppub}
\usepackage{graphicx}
\usepackage{mathtools}
\usepackage{array}
\usepackage{amsfonts}
\usepackage[utf8]{inputenc} 
\usepackage{amsmath}
\usepackage{pdfpages}
\usepackage{hepunits}
\usepackage{physics}
\usepackage{bm}
\usepackage{tikz}
\usepackage{dsfont}
\usepackage[english]{babel}

\usepackage{verbatim}
\usepackage[thinlines]{easytable}
\usepackage{dsfont}

\DeclarePairedDelimiter\floor{\lfloor}{\rfloor}

\title{Chaos in CFT dual to rotating BTZ}
\author{Ben Craps, Surbhi Khetrapal, Charles Rabideau}
\affiliation{Theoretische Natuurkunde, Vrije Universiteit Brussel (VUB) and
	The International Solvay Institutes, B-1050 Brussels, Belgium}
\emailAdd{ben.craps@vub.be, surbhi.khetrapal@vub.be, charles.rabideau@gmail.com}
\abstract{We compute out-of-time-order correlators (OTOCs) in two-dimensional holographic conformal field theories (CFTs) with different left- and right-moving temperatures. Depending on whether the CFT lives on a spatial line or circle, the dual bulk geometry is a boosted BTZ black brane or a rotating BTZ black hole. In the case when the spatial direction is non-compact, we generalise a computation of Roberts and Stanford and show that to reproduce the correct bulk answer a maximal channel contribution needs to be selected when using the identity block approximation. We use the correspondence between global conformal blocks and geodesic Witten diagrams to extend our results to CFTs on a spatial circle. 

In \cite{Mezei:2019dfv} it was shown that the OTOC for a rotating BTZ black hole exhibits a periodic modulation about an average exponential decay with Lyapunov exponent $2\pi/\beta$. In the extremal limit where the black hole is maximally rotating, it was shown in \cite{Craps:2020ahu} that the OTOC exhibits an average cubic growth, on which is superposed a sawtooth pattern which has small periods of Lyapunov growth due to the non-zero temperature of left-movers in the dual CFT. Our computations explain these results from a dual CFT perspective.}

\begin{document}
\maketitle

\section{Introduction}

Quantum chaos has emerged as an essential tool to study real time dynamics of quantum many body systems. In recent years, out-of-time-order correlators (OTOCs) have been used to probe the dynamical chaotic behaviour in quantum field theories and the gravitational systems which are holographically dual to them \cite{Shenker:2013pqa,Maldacena:2015waa,Shenker:2013yza,Shenker:2014cwa,Roberts:2014ifa,Roberts:2014isa, Kitaev:2015}. The OTOC was shown to exhibit a phase of exponential decay for theories which admit a holographic bulk dual. 

In \cite{Mezei:2019dfv}, it was shown that the OTOC in a rotating BTZ black hole exhibits an average exponential decay on which is superposed a sawtooth pattern describing alternating exponential behaviours determined by the left and right moving temperatures of the dual CFT. A sawtooth pattern was also observed in the extremal limit of the rotating BTZ black hole \cite{Craps:2020ahu}. Thus it is natural to wonder how this behaviour emerges in the CFT holographically dual to rotating BTZ.

In this work we will therefore study the OTOC in rotating and boosted ensembles in 2d CFT. We start with putting the CFT on a line, where a boosted ensemble is dual to a black brane background. Much of the computation of the OTOC in a boosted ensemble appeared previously in \cite{Halder:2019ric}, except for an important subtlety that affects the final result. The computation in the dual black brane geometry was performed in \cite{Poojary:2018esz,Jahnke:2019gxr,Mezei:2019dfv}. In \cite{Mezei:2019dfv}, it was found that at late times the smaller of the left and right moving temperatures controlled the Lyapunov exponent. It was also remarked that a prescription of maximisation over the different channels in which the conformal block decomposition can be performed is required and might provide a mechanism in the CFT for this behaviour. We study these channels in detail, perform the required maximisation and find that this indeed reproduces the computation in the black brane. Since the Lyapunov exponent is associated with the decay of the OTOC, within the Lyapunov regime the prescription of maximising over channels ensures that the lower of the two temperatures sets the Lyapunov exponent at late times. This is discussed in detail in section \ref{sec:line}.

With this result for boosted ensembles on the line in hand, we next analyse the case of rotating ensembles on the circle. Gravitational computations of the OTOC have found that in the Lyapunov regime the instantaneous Lyapunov exponent exhibits a sawtooth pattern which when averaged over a full period gives an averaged Lyapunov exponent controlled by the temperature \cite{Mezei:2019dfv,Craps:2020ahu}. We show that in the Lyapunov regime, where the global conformal block of the stress tensor dominates, the OTOC on the circle can be obtained by that on the line from a reasoning that involves the relation between global conformal blocks and geodesic Witten diagrams. In this way we reproduce the gravitational results found in \cite{Mezei:2019dfv}. We also study the extremal and zero-temperature limits, which require backing off from some of the approximations used in the previous result, and reproduce the gravitational results found in \cite{Craps:2020ahu}. 

The paper is organised as follows: In section~\ref{sec:BTZ}, we briefly review the rotating BTZ black hole and its dual CFT. In section~\ref{sec:line}, we compute the OTOC in the thermal CFT with non-compact spatial direction dual to a boosted black-brane in the bulk. In section~\ref{sec:circle}, we use geodesic Witten diagrams to develop a method of images to approximate the $4$-point function on the torus, which we then use to compute the OTOC in a thermal CFT with compact spatial direction dual to rotating BTZ black hole in the bulk. In section~\ref{sec:extremal}, we compute the OTOC in the extremal limit. Appendix~\ref{app:vacuum} contains a computation of the OTOC in the vacuum of a 2d CFT on a spatial circle.

\section{Rotating BTZ black hole}\label{sec:BTZ}
In this section, we briefly review the rotating BTZ black hole \cite{Banados:1992wn,Banados:1992gq}, whose metric in Schwarzschild co-ordinates is
\begin{align}\label{rotBTZbh}
ds^2 & = -f(r)dt^2 + \frac{dr^2}{f(r)} + r^2 \left(d\varphi - \frac{r_+r_-}{ r^2} dt\right)^2 \nonumber\\
f(r) & = \frac{(r^2-r_+^2)(r^2-r_-^2)}{ r^2},
\end{align}
where the AdS radius has been set to $1$ and $\varphi$ is the angular co-ordinate. When the angular co-ordinate is periodic, $\varphi \sim \varphi + 2\pi$, the above metric describes a black hole and when $\varphi$ is non-periodic, it describes a black brane geometry. The horizon radii $r_\pm$ are related to the mass $M$, angular momentum $J$, Hawking temperature $T$ and angular potential $\Omega$ of the black hole,
\begin{align} \label{rotBTZconserved}
M = r_+^2+r_-^2, \quad J = 2r_+r_-, \quad T=\frac{r_+^2-r_-^2}{2\pi r_+}, \quad \Omega=\frac{r_-}{r_+},
\end{align}
and the angular potential takes values $0 \leq \Omega\leq 1$. In the maximally rotating limit, $r_+ = r_-$ and $\Omega =1$. Continuing to a Euclidean black hole would involve taking $t$, $r_-$, $J$ and $\Omega$ to be purely imaginary.

Let us also introduce co-moving co-ordinates $(t,r,\phi)$, where 
\begin{align} \label{rotBTZphi}
\phi = \varphi - \Omega \, t,
\end{align}
in terms of which the metric takes the form
\begin{align}
ds^2 = -f(r)dt^2 + \frac{dr^2}{f(r)}+r^2 \left(\frac{r_-}{r_+}\frac{r^2-r_+^2}{r^2}dt+d\phi\right)^2.
\end{align}
In these co-ordinates, the angular velocity at the horizon vanishes, $\Omega_H=0$. 

\subsection{CFT dual to rotating BTZ}
The CFT dual to rotating BTZ has unequal left- and right-moving temperatures and its partition function is given by \cite{Caputa:2013eka}
\begin{align}
Z(\beta_L,\beta_R) = \text{Tr} \, e^{-\beta_L E_L - \beta_R E_R}.
\end{align}
Here, the inverse temperatures of left- and right-movers are given by
\begin{align}
\beta_L=\beta(1-\Omega), \qquad \beta_R=\beta(1+\Omega)
\end{align}
and are written in terms of inverse temperature $\beta$ and the chemical potential for conserved angular momentum, $\Omega$. In Lorentzian signature, the space-time co-ordinates on the 2d CFT are $(\sigma,t)$, where $\sigma = \ell \varphi$, and we redefine $t_{\rm new}=\ell \, t_{\rm old}$. In Euclidean signature, we introduce complex co-ordinates  $x = \sigma + i \tau, \bar{x} = \sigma -i \tau$, where the Euclidean time $\tau$ is related to the Lorentzian time $t$ by $\tau = i t$. If $\varphi$ is periodic, i.e.\ dual to a black hole in the bulk, the 2d thermal CFT has compact spatial direction and hence is on a torus. If $\varphi$ is non-periodic, i.e.\ dual to black brane geometry in the bulk, the 2d CFT is on a thermal cylinder. The thermal cylinder can be mapped to a complex plane $(z,\bar{z})$ by
\begin{align} \label{cyl_to_plane_map}
z=e^{\frac{2\pi}{\beta_L}x}, \qquad \bar{z}=e^{\frac{2\pi}{\beta_R}\bar{x}}.
\end{align} 
Note that, since in Euclidean co-ordinates the chemical potential for conserved angular momentum is imaginary,
\begin{align}
    \Omega = i \, \Omega_E,
\end{align}
the co-ordinates $z$ and $\bar{z}$ in \eqref{cyl_to_plane_map} are complex conjugate in Euclidean signature.

%%%%%%%%%%%%%%%%%%%%%%%%%%%%%%%%%%%%%%%%%%%%%%%%%%%%%%%%%%%%%%%
\section{OTOC for 2d CFT on a line} 
\label{sec:line}

In this section, we compute the OTOC in a two-dimensional thermal CFT with a non-zero angular potential of rotation, where the spatial direction is non-compact. When viewed as a limit of a CFT on a spatial circle with periodic $\varphi$, it could be obtained by introducing $\sigma=\ell \varphi$ and working in the regime where $\ell \to \infty$ while keeping $\sigma$ finite. Our starting point is the following 4-point function on the Euclidean thermal cylinder with different temperatures for left- and right-movers:
\begin{align}
\langle W(x_1,\bar{x}_1) W(x_2,\bar{x}_2) V(x_3,\bar{x}_3) V(x_4,\bar{x}_4)\rangle_{\beta_L,\beta_R},
\end{align}
where
\begin{align} \label{cylinder_coord}
x_i = \sigma_i + i \tau_i, \qquad \bar{x}_i = \sigma_i - i \tau_i
\end{align}
are the holomorphic and anti-holomorphic co-ordinates on the cylinder. As before, the Euclidean time is related to Lorentzian time by $\tau_i = i \, t_i$. The Euclidean correlator is single-valued. Continuing to Lorentzian signature introduces multivaluedness, which reflects a dependence on operator ordering. The prescription used in \cite{Roberts:2014ifa} to obtain a specific operator ordering from the Euclidean correlator is to assign small imaginary time to each operator: $\tau_j =  \epsilon_j$ $(t_i \to t_i-i \epsilon_i)$; then, with imaginary times held fixed, the real times are increased to their required Lorentzian value. Lastly, the operators are smeared in real time and the imaginary times are taken to be zero. (As in \cite{Roberts:2014ifa}, we will leave this last step implicit.)
When the small parameters satisfy the relation $\epsilon_1<\epsilon_3<\epsilon_2<\epsilon_4$, the correlator gives the required OTOC. 

The location of the operators in the $(\sigma,t)$ plane corresponds to
\begin{align}
{\cal C} (\sigma, t) = \frac{\langle  W(0,-i\epsilon_1) V(\sigma,t-i\epsilon_3)  W(0,-i\epsilon_2)V(\sigma,t-i\epsilon_4) \rangle}{\langle W(0,-i\epsilon_1) W(0,-i\epsilon_2) \rangle \langle V(\sigma,t-i\epsilon_3) V(\sigma,t-i\epsilon_4) \rangle},
\end{align}
where we have normalized the OTOC by dividing by a product of 2-point functions.
Using the cylinder-to-plane map \eqref{cyl_to_plane_map}, the location of the operators in the $(z,\bar{z})$ plane is 
\begin{align} \label{OTOC_loc}
z_1 & = e^{\frac{2\pi}{\beta_L}i\epsilon_1} \qquad \qquad \qquad \,\, \bar{z}_1 = e^{-\frac{2\pi}{\beta_R}i\epsilon_1} \nonumber\\
z_2 & = e^{\frac{2\pi}{\beta_L}i\epsilon_2} \qquad \qquad \qquad \,\, \bar{z}_2 = e^{-\frac{2\pi}{\beta_R}i\epsilon_2} \nonumber\\
z_3 & = e^{\frac{2\pi}{\beta_L}\left(\sigma-t+i\epsilon_3\right)} \qquad \qquad \bar{z}_3 = e^{\frac{2\pi}{\beta_R}\left(\sigma+t-i\epsilon_3\right)} \nonumber\\
z_4 & = e^{\frac{2\pi}{\beta_L}\left(\sigma-t+i\epsilon_4\right)} \qquad \qquad \bar{z}_4 = e^{\frac{2\pi}{\beta_R}\left(\sigma+t-i\epsilon_4\right)}.
\end{align}
Using the Virasoro block decomposition of 4-point correlation functions, the OTOC is given by
\begin{align}
{\cal C}(\sigma,t) = \sum_{h,\bar{h}} P_{h,\bar{h}}{\cal V}_{h,\bar{h}}(u,v).
\end{align}
Here, ${\cal V}_{h,\bar{h}}$ are the Virasoro blocks, $P_{h,\bar{h}}$ are the  theory-dependent Virasoro block coefficients, and $u,v$ are functions of the conformal cross-ratios,
\begin{align}
u = \eta \, \bar{\eta} , \qquad v=\left( 1-\eta \right)\left( 1- \bar{\eta} \right).
\end{align}
The conformal cross-ratios are
\begin{align} \label{eq:cross-ratio_def}
\eta = \frac{z_{12}z_{34}}{z_{13}z_{24}}, \qquad \bar{\eta} = \frac{\bar{z}_{12} \bar{z}_{34}}{\bar{z}_{13} \bar{z}_{24}},
\end{align}
where $z_{ij} = z_i-z_j$ and $\bar{z}_{ij} = \bar{z}_i-\bar{z}_j$. Substituting the location of the operators from \eqref{OTOC_loc}, the cross-ratios become
\begin{align} \label{eqn:cross-ratio}
\eta = \frac{-\sin \left( \frac{\pi}{\beta_L}\epsilon_{12} \right) \sin \left( \frac{\pi}{\beta_L}\epsilon_{34} \right)}{\sinh \left(  \frac{\pi}{\beta_L}\left(t- \sigma +i\epsilon_{13} \right)\right) \sinh \left(  \frac{\pi}{\beta_L}\left(t- \sigma +i\epsilon_{24} \right)\right)}\\
\bar{\eta} = \frac{-\sin \left( \frac{\pi}{\beta_R}\epsilon_{12} \right) \sin \left( \frac{\pi}{\beta_R}\epsilon_{34} \right)}{\sinh \left(  \frac{\pi}{\beta_R}\left(t + \sigma +i\epsilon_{13} \right)\right) \sinh \left(  \frac{\pi}{\beta_R}\left(t + \sigma +i\epsilon_{24} \right)\right)}.
\end{align}

\begin{figure}
\centering
\includegraphics[scale=1]{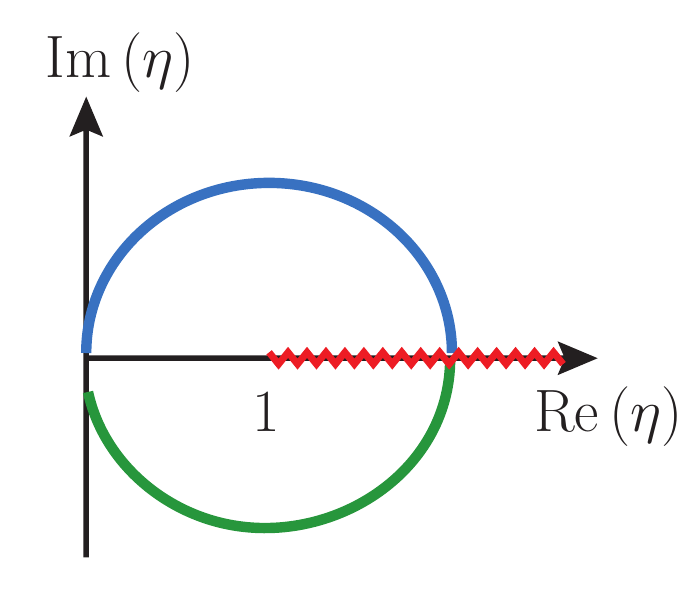}
\caption{The blue curve denotes $t< \sigma$, at $t=\sigma$ the holomorphic cross-ratio crosses branch cut running from $\eta=1$ to $\infty$ and enters the second sheet where the green curve denotes $t>\sigma$ and $\eta$ becomes small as $t \gg \sigma$.}
\label{fig1}
\end{figure}

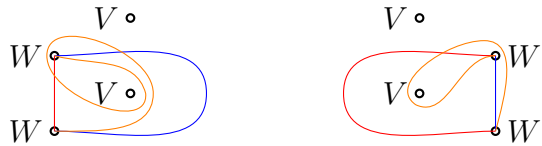
\begin{figure}
    \centering
    \begin{tikzpicture}
    \draw[thick] (-1,0) circle (.05) node[left]{$W$} ;
    \draw[thick] (-1,-1) circle (.05) node[left]{$W$} ;
    \draw[thick] (0,.5) circle (.05) node[left]{$V$} ;
    \draw[thick] (0,-.5) circle (.05) node[left]{$V$} ;
    \draw[red] (-1,0) -- (-1,-1);
    \draw[blue] (-1,0) to [out=0,in=90] (1,-.5) to [out=-90,in=0] (-1,-1);
    \draw[orange] (-1,0) to [out=-15,in=105] (.2,-.5) to [out=180+95,in=-90] (-1.1,0) to [out=90,in=90] (.3,-.5) to [out=-90,in=0] (-1,-1);
    \end{tikzpicture}
    \qquad\qquad 
 \begin{tikzpicture}
    \draw[thick] (1,0) circle (.05) node[right]{$W$} ;
    \draw[thick] (1,-1) circle (.05) node[right]{$W$} ;
    \draw[thick] (0,.5) circle (.05) node[left]{$V$} ;
    \draw[thick] (0,-.5) circle (.05) node[left]{$V$} ;
    \draw[blue] (1,0) -- (1,-1);
    \draw[red] (1,0) to [out=180,in=90] (-1,-.5) to [out=-90,in=180] (1,-1);
    \draw[orange] (1,0) to [out=180,in=-25] (-.05,-.65) to [in=180,out=145] (.85,.2) to [out=0,in=75] (1,-1);
    \end{tikzpicture}
    \caption{Different channels corresponding to paths along which the pair of $W$ operators can be contracted. The colours denote channels that should be identified across the two diagrams if the $W$ operators are moved horizontally to the right. However, if the $W$ operators where moved below the $V$ operators the red path in the left diagram would be identified with the blue path in the right diagram.
    This leads to the branch cut in the conformal block. }
    \label{fig:channels}
\end{figure}

Next, we use the Virasoro identity block approximation, which approximates the correlator by the contribution of the Virasoro block of the identity operator, ${\cal V}_{0,0}(u,v)$. The Virasoro blocks factorise into functions of purely holomorphic and anti-holomorphic components,
\begin{align} \label{virasoro_block_factorise}
{\cal V}_{0,0}(u,v) = {\cal V}_0(\eta) {\cal V}_0(\bar{\eta}).
\end{align}  
Even in Euclidean signature, the blocks, including $\mathcal{V}_0(\eta)$, are multivalued due to a branch cut in the complex $\eta$ plane from $1$ to $\infty$ (see figure \ref{fig1}). 
The specification of the OPE channel used to derive the block expansion starts with specifying which pairs of operators are contracted. The branch cut is a reflection of the fact that the block has a monodromy when one of the operators from one pair is brought around one of the operators in the other pair. This means that the specification of a channel requires, in addition to a choice of which operators to contract, a choice of path along which to contract them,\footnote{This can also be thought of in terms of a choice of a region homotopic to a disk by slightly blowing up this path. The validity of the OPE expansion is ensured by a conformal map of this region to the unit disk followed by a contraction of the unit disk to bring the pairs of operators together.} as depicted in figure \ref{fig:channels}. The choice of which sheet of the multivalued function $\mathcal{V}_0(\eta)$ to use corresponds to this choice of path along which the operators are contracted.
Under the assumption that there is a choice of channel where the vacuum block provides a good approximation to the correlator, it must be the channel where the vacuum block provides the largest contribution \cite{Asplund:2014coa}. Note that fixing the path along which the operators are to be contracted fixes the channel for both the holomorphic and anti-holomorphic blocks, so that there is not two independent choices of channel. This means that along the Euclidean section, where $\bar \eta = \eta^*$, the two blocks are on opposite sheets.\footnote{When $\eta$ crosses the branch cut in one direction, $\bar \eta$ crosses it in the other way so that the two functions end up on opposite sheets. That is when $1-\eta \rightarrow e^{ -2 \pi i k}(1-\eta)$, then $1-\eta^* \rightarrow e^{ 2 \pi i k}(1-\eta^*)$.} This choice is labelled by a single integer, which we will denote $k$.

Once the block is continued to Lorentzian times, the cross-ratios can cross branch cuts and the two blocks may end up on different sheets. As explained in \cite{Roberts:2014ifa}, the cross-ratio crosses the branch cut in the block along the light cone of the OTOC. 
\begin{figure}
    \centering
\begin{tikzpicture} 
\draw[->, thick] (-2,0)--(2,0) node[right]{$\sigma$};
\draw[->, thick] (0,-2)--(0,2) node[above]{$t$};
\draw [dashed] (-2,-2)--(2,2) node[right]{\footnotesize $t=\sigma$};
\draw [dashed] (2,-2)--(-2,2) node[right]{\footnotesize $t=-\sigma$};
\draw[->,red] (1,0) to [out=75,in=330] (.35,1.5);
\draw[->,blue] (-1,0) to [out=90,in=195] (.25,1.5);
 \draw[thin] (.25,1.55) -- (.35,1.45);
 \draw[thin] (.25,1.45) -- (.35,1.55);
\end{tikzpicture}
    \caption{The location of the branch cuts (dashed lines) in the conformal block in the $\sigma$-$t$ plane. There is a branch cut in the holomorphic block at $t=\sigma$ and in the anti-holomorphic block at $t=-\sigma$. When continuing the correlator from the $t=0$ Euclidean slice into the future timelike region, one or the other of these branch cuts must be crossed leading to two candidate channels. These correspond to the continuations along the red or blue arrows. Note that along the Euclidean section the two branch cuts coincide, ensuring that in the spacelike region the holomorphic and anti-holomorphic blocks are on opposite sheets. }
    \label{fig:cuts}
\end{figure}
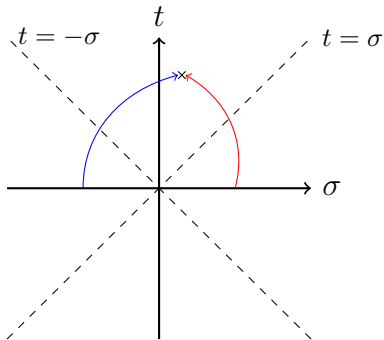
Indeed, for $|t-\sigma| \gg \epsilon$, the cross-ratio $\eta$ sits just above or below for $t-\sigma$ negative or positive, respectively. When $t=\sigma$, 
\begin{align}
    \eta \sim \frac{\epsilon_{12} \epsilon_{34}}{\epsilon_{13}\epsilon_{24} }\,, \qquad \qquad (t=\sigma)\,. \label{eqn:lightcone_cross-ratio}
\end{align}
This is greater than 1 for the OTOC, ensuring that the branch cut in the holomorphic block is crossed when the $t=\sigma$ line is crossed. Similarly, the $t=-\sigma$ line corresponds to the branch cut in the anti-holomorphic block. For $t=0$ we return to the Euclidean section where the branch cut for both blocks occurs at the same location. The location of the branch cuts in the $\sigma$-$t$ plane is summarised in figure~\ref{fig:cuts}. 
For time-ordered correlators, the cross-ratio at the lightcone, \eqref{eqn:lightcone_cross-ratio}, is less than 1, so that it does not cross the branch cut.

In order to compute the OTOC, we will maximise over all possible channels, or in other words over sheets of the multivalued block. We therefore turn to the explicit form of the identity block in the semiclassical limit described in \cite{Roberts:2014ifa,Fitzpatrick:2014vua},
\begin{align} \label{eqn:semi-classical_block}
{\cal V}_0(\eta) \approx \left( \frac{\eta}{1- \left( 1-\eta \right)^{1-12h_w/c}} \right)^{2h_v} \,,
\end{align}
valid for a CFT with large central charge $c$ such that  $\frac{h_w}{c}$ is held fixed and small and $h_v$ satisfies, $1 \ll h_v \ll c$. In particular $h_v$ does not scale as $c$. 
This has the expected branch cut along $[1,\infty)$.

When $t - \sigma\gg \beta_{L}$ and $t + \sigma\gg \beta_{R}$,
the cross-ratios become
\begin{align} \label{cross-ratio-large-temp}
\eta \approx -e^{-\frac{2\pi}{\beta_L}(t-\sigma)} {\tilde{\epsilon}}^{L*}_{12} \,{\tilde{\epsilon}}^L_{34}, \qquad \bar{\eta} \approx -e^{-\frac{2\pi}{\beta_R}(t+\sigma)} {\tilde{\epsilon}}^{R*}_{12} \,{\tilde{\epsilon}}^R_{34} ,
\end{align}
where
\begin{align}
{\tilde{\epsilon}}^X_{ij} = i \left( e^{\frac{2\pi}{\beta_X} i \epsilon_i} - e^{\frac{2\pi}{\beta_X}i \epsilon_j}\right), \quad X=L,R.
\end{align}
Therefore, the behaviour of the conformal block near $\eta \sim0$ is relevant for understanding the OTOC for large times. 
On the $k^\mathrm{th}$ sheet, where we send 
\begin{align}\label{eq:eta_sheet_k}
\left( 1- \eta \right) \to \left( 1- \eta \right)e^{-2\pi i k}\,,
\end{align}
near $\eta \sim0$ the approximation \eqref{eqn:semi-classical_block} behaves as
\begin{align} \label{vac_block_eta_0}
{\cal V}_0(\eta) 
\sim \left(-\frac{24\pi i k  h_w }{c \eta} + 1-\frac{12 h_w (1-2 \pi i k)}{c}  + O(\eta) + O\left( \left(h_w/c\right)^2\right)  \right)^{-2h_v} \,.
\end{align}
At sufficiently small $\eta$, that is at late times, this correlator goes as $\eta^{2h_v}$, which are the expected late-time Ruelle resonances dual to quasi-normal mode decay. 
We will instead be interested in the intermediate regime where Lyapunov growth is known to occur, $\frac{h_v h_w}{c \eta} \ll 1$,
\begin{align}
    {\cal V}_0(\eta) &\approx \left(1 -  \frac{24\pi i k  h_w }{c \eta}  \right)^{-2h_v}\nonumber\\
&\approx 1 + \frac{48\pi i k h_v h_w }{c \eta} - \frac12 \left(\frac{48\pi i k h_v h_w }{c \eta}\right)^2 +O\left(h_v^{-1}, \left(\frac{h_v h_w}{c \eta}\right) ^3 \right)\,.
\end{align} 
Recall that the OTOC is the product of the holomorphic and anti-holomorphic blocks,
\begin{align} \label{eqn:OTOC_channels}
    \mathcal{C}(\sigma,t) \approx \left(1 -  \frac{24\pi i k  h_w }{c \eta}  \right)^{-2h_v}  \left(1 -  \frac{24\pi i \bar k  h_w }{c \bar \eta}  \right)^{-2h_v} \,,
\end{align}
where $k$ denotes the sheet of ${\cal V}_0(\eta)$ and $\bar k$ the sheet of ${\cal V}_0(\bar \eta)$. On the Euclidean section these are opposite, $\bar k = - k$.
However, as described above, when we continue to Lorentzian time, crossing the $t=\pm \sigma$ lightcones in the direction of increasing time leads to $k\rightarrow k+1$ or $\bar k \rightarrow \bar k +1$ respectively. In either case, for times in the future timelike region, $\bar k = -k+1$.

Therefore, in the regions where the OTOC is spacelike and no lightcones have been crossed, our prescription is that \eqref{eqn:OTOC_channels} must be maximised with $\bar k = -k$. This happens for $k=0$, where $\mathcal{C} \approx 1$. When we cross the lightcone into the future timelike region, we instead have $\bar k = -k+1$. Clearly increasing $|k|$ or $|\bar k|$ causes \eqref{eqn:OTOC_channels} to decrease and so there are only two candidates that need to be compared in detail: $(k=1,\bar k =0)$ and $(k=0,\bar k=1)$. These correspond to 
\begin{align} \label{eqn:OTOC_candidate_channels}
    \mathcal{C}(\sigma,t) \approx \left(1 -  \frac{24\pi i  h_w }{c \eta}  \right)^{-2h_v} \quad \mathrm{or}\quad  \left(1 -  \frac{24\pi i  h_w }{c \bar \eta}  \right)^{-2h_v} \,,
\end{align}
respectively. These two channels could have been obtained by thinking of continuing the channel that is dominant for $\sigma>0$ or $\sigma<0$, respectively, along increasing times into the future lightcone, but we emphasise that our prescription was to maximise over all channels in the final result. These two channels compete in the future lightcone and whichever is dominant at a particular value of $\sigma$ and $t$ must be chosen. This guarantees that the answer is continuous, although it will not be smooth at the transition between the two channels. This is a consequence of the large $c$ limit. Since the commutator squared is $2(1-\mathrm{Re}\,\mathcal{C})$, the maximisation over channels ensures that, for late enough times still within the range of validity of the Lyapunov regime, the slower of the two potential Lyapunov exponents will be realised. 

The exchange of dominance between the two candidate channels occurs when $\eta = \bar \eta$, which happens when 
\begin{align}\label{sigmaomegat}
   \sigma = \Omega \, t -\frac{\beta (1-\Omega^2)}{2\pi}  \log \left( \frac{1+\Omega}{1-\Omega} \right) .
\end{align}

Substituting \eqref{cross-ratio-large-temp} in \eqref{eqn:OTOC_candidate_channels} and choosing the maximal channel, for $t>|\sigma|$ we find the OTOC to be
\begin{align} \label{otoc_line_hhll}
{\cal C}(\sigma,t) & \approx \left( 1+\frac{24\pi i h_w}{{\tilde{\epsilon}}^{L*}_{12} {\tilde{\epsilon}}^L_{34} c} \exp \left(\frac{2\pi}{\beta_L}(t - \sigma)\right) \right)^{-2h_v}, \;\;   t < \frac{\sigma^*}{\Omega}\\
 & \approx \left( 1+\frac{24\pi i h_w}{{\tilde{\epsilon}}^{R*}_{12} {\tilde{\epsilon}}^R_{34} c} \exp \left(\frac{2\pi}{\beta_R}(t + \sigma)\right) \right)^{-2h_v}, \;\; t \geq  \frac{\sigma^*}{\Omega} ,
\end{align}
where we have defined $\sigma^*$ as
\begin{align}\label{sigma_star}
    \sigma^* \equiv \sigma + \frac{\beta (1-\Omega^2)}{2\pi }  \log \left( \frac{1+\Omega}{1-\Omega} \right).
\end{align}
Further, in the regime where $\frac{h_w h_v}{c\, {\tilde{\epsilon}}^{L,R*}_{12} {\tilde{\epsilon}}^{L,R}_{34} } e^{\frac{2\pi}{\beta_{L,R}}(t\mp \sigma)} \ll1$,
\begin{align} \label{OTOC_line_beta}
{\cal C}(\sigma,t) & \approx 1 - \frac{48 \pi i h_w h_v}{c \, {\tilde{\epsilon}}^{L*}_{12} {\tilde{\epsilon}}^L_{34} }
e^{\frac{2\pi}{\beta_L}(t-\sigma)}, \qquad t < \frac{\sigma^*}{\Omega}\\
& \approx 1 - \frac{48 \pi i h_w h_v}{c \,{\tilde{\epsilon}}^{R*}_{12} {\tilde{\epsilon}}^R_{34}} e^{\frac{2\pi}{\beta_R}(t+\sigma)}, \qquad t \geq \frac{\sigma^*}{\Omega} .
\end{align} 
For small $\epsilon_i$, we can re-write the above equation in the form
\begin{align}\label{OTOC_line_beta1}
    {\cal C}(\sigma,t)  \approx 1- \frac{48 i h_w h_v \beta^2}{c\, \pi \,\epsilon_{12} \epsilon_{34}}(1+\Omega)^{1+\Omega}(1-\Omega)^{1-\Omega} \begin{cases}
    e^{\frac{2\pi}{\beta_L}(t-\sigma^*)},& t<\frac{\sigma^*}{\Omega} \\
    e^{\frac{2\pi}{\beta_R}(t+\sigma^*)} & t \geq \frac{\sigma^*}{\Omega}.
    \end{cases}
\end{align}
For $\Omega >0$, at late times the decay of the OTOC is governed by the Lyapunov exponent $\lambda_R = \frac{2\pi}{\beta_R} < \frac{2\pi}{\beta}$. This is in accordance with the result from bulk computations in \cite{Mezei:2019dfv} that the smaller Lyapunov exponent governs the decay of the OTOC at late times and was guaranteed by the maximisation over channels as explained above. 
Comparing to the OTOC derived from bulk computations in a decompactification limit of the rotating BTZ black hole in \cite{Mezei:2019dfv}, our result contains cutoff- and $\Omega$-dependent prefactors which \cite{Mezei:2019dfv} did not keep track of. In addition, it has $\sigma^*$ appearing in place of $\sigma$, which corresponds to a shift in time.

\begin{figure}
    \centering
    \includegraphics[scale=0.3]{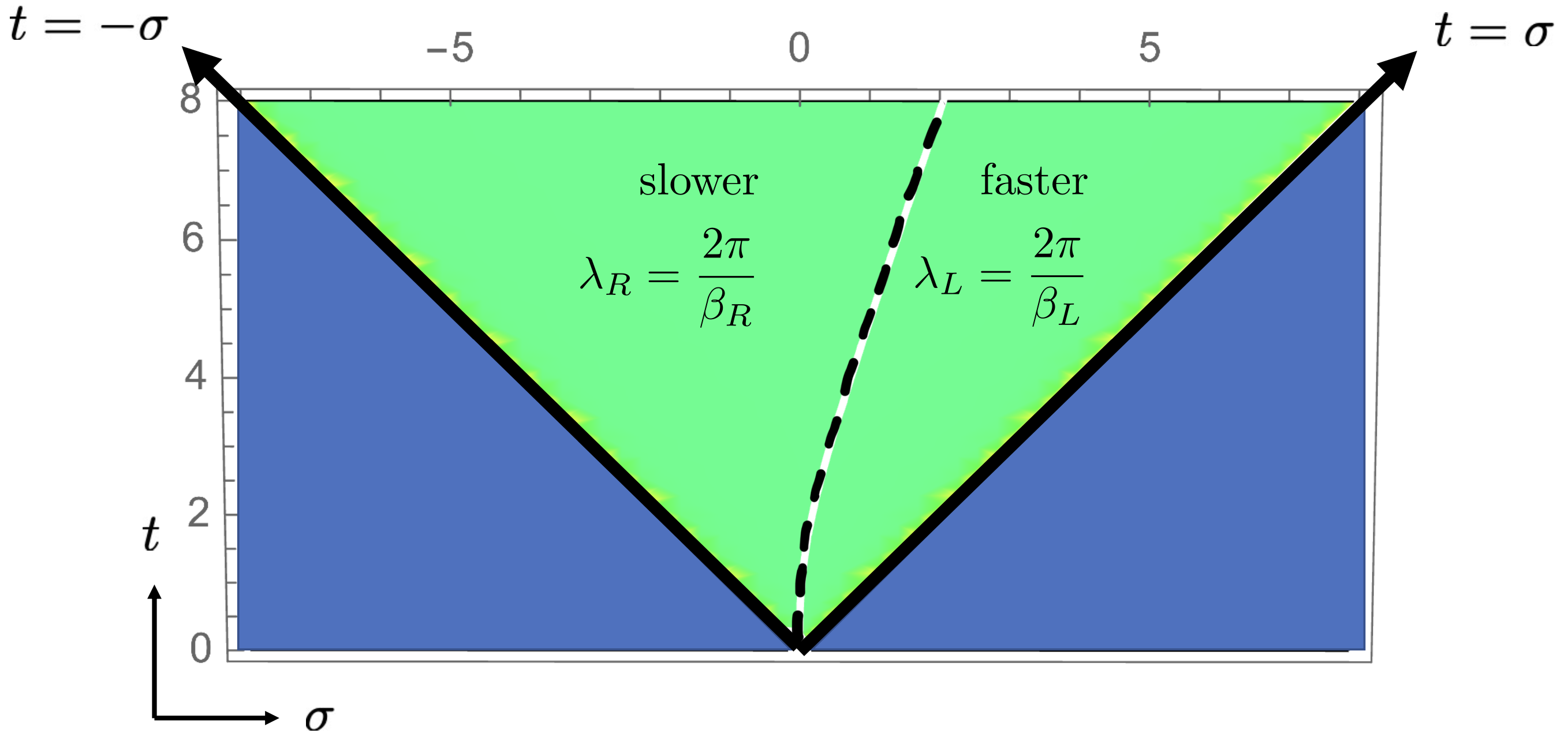}
    \caption{Regions with the two Lyapunov growths separated by the black dashed line, which for early times intersects the origin and for late times asymptotes to $t=\frac{\sigma^*}{\Omega}$. The plot was obtained numerically for the following values of parameters: $\beta = 2\pi, \Omega = \frac{1}{3}, c=10, h_w = 5$ and $h_v=1.5$. For negative $\sigma$, in conventions where $\Omega>0$, the slower Lyapunov exponent determines the growth with time in the future lightcone. For positive $\sigma$, one encounters first the faster and then the slower Lyapunov exponent.}
    \label{fig:lightcone}
\end{figure}

However, the expression for this shift in \eqref{sigmaomegat} is only valid in the late-time regime $t \gg \beta_{L,R}$, since equations \eqref{sigmaomegat} to \eqref{OTOC_line_beta1} were obtained in this approximation, as can be seen from \eqref{cross-ratio-large-temp} and \eqref{vac_block_eta_0}. In Figure~\ref{fig:lightcone}, we determine the values of $(t,\sigma)$ at which the exchange of dominance between the two channels takes place, without using this late-time approximation. We obtain the OTOC numerically by substituting equations \eqref{eqn:cross-ratio}, \eqref{eqn:semi-classical_block} and \eqref{eq:eta_sheet_k} in \eqref{virasoro_block_factorise} for the following values of the parameters: $\beta = 2\pi, \Omega = \frac{1}{3}, c=10, h_w = 5$ and $h_v=1.5$ (with $\epsilon_1 = 0.01,\ \epsilon_2=0.03,\ \epsilon_3 = 0.02,\ \epsilon_4 = 0.04$). We find that at early times the line separating the two channels passes through the origin and at late times it asymptotes to \eqref{sigmaomegat}.

Alternatively, one could regulate the OTOC with Lorentz scalars instead of vectors by choosing $\epsilon_{L,R} = \beta_{L,R} \, x_i$, where $x_i$ are Lorentz invariant regulators (corresponding to moving operator insertions a fraction $x_i$ of the thermal circle). Using these regulators, we find that the exchange of dominance between the two channels happens at $t = {\sigma}/{\Omega}$, thus the line separating the slow and fast growth regions in Figure~\ref{fig:lightcone} is a straight line passing through the origin.\footnote{We thank the anonymous referee for questions about the cutoff dependence, and in particular for suggesting this alternative regulator.}

We wish to contrast the prescription we have used with that appearing in previous works on CFT computations of OTOCs. In particular, in the seminal work of \cite{Roberts:2014ifa}, the authors seem to have only considered the $k=\bar{k}=0$ channel in the space-like region with $\sigma>0$, which they then continued to the time-like region by increasing $t$ while keeping $\sigma$ fixed. This does indeed produce the correct result in the non-rotating case they considered. Taking this prescription at face value led to some confusion in the rotating case \cite{Halder:2019ric}, where a result for the OTOC was found which was discontinuous at $\sigma =0$ and did not agree with the result obtained from boosting the result in the thermal ensemble \cite{Mezei:2019dfv}. 
These issues 
can be remedied by maximising over channels, as was suggested in \cite{Mezei:2019dfv}. In this work, we have demonstrated in detail that this is indeed the case. 

We also would like to point out that only the slower of the two potential Lyapunov exponents is realised in the OTOC at late time. This is in contrast with the na\"ive expectation, based on intuition from classical mechanics, that the largest Lyapunov exponent always dominates. Of course in a system with different modes growing exponentially the largest growth will always eventually dominate. However, in this case the potential Lyapunov exponents, which were also identified in \cite{Poojary:2018esz}, correspond to different channels of the block expansion and only the slower one is realised in the physical growth of the commutator squared. This matches the gravitational result found in \cite{Mezei:2019dfv}.

%%%%%%%%%%%%%%%%%%%%%%%%%%%%%%%%%%%%%%%%%%%%%%%%%%%%%%%%%%%%%%%%%%%%%%%%%%%%%%%%%%%%%%%%%%%%%%%%%%%%%%%%%%%%%%%%%%%%
\section{OTOC for 2d CFT on a spatial circle}\label{sec:circle}

In this section we describe how to extend some of our results for a CFT on a line to a CFT on a spatial circle. Specifically, we will make use of the relation between global conformal blocks and geodesic Witten diagrams to extract a prescription for computing OTOCs for CFTs on a line, which we then put to use. Finally, we comment on a subtlety related to the fact that black holes do not dominate the canonical ensemble at low temperatures.

\subsection{Conformal block as geodesic Witten diagram}
In a CFT, a 4-point function can be written as an expansion in global conformal blocks,
\begin{align} \label{block_expansion}
    \langle  W(z_1,\bar{z}_1) W(z_2,\bar{z}_2) V(z_3,\bar{z}_3) V(z_4,\bar{z}_4) \rangle  = \frac{1}{|z_{12}|^{4h_w}|z_{34}|^{4h_v}} \sum_{{\cal O}} \lambda_{WW {\cal O}} \lambda_{VV {\cal O}} G_{h,\bar{h}}(u,v),
\end{align}
where the exchanged primary ${\cal O}$ is an operator of holomorphic and anti-holomorphic dimensions $(h,\bar{h})$. Here, $(u, v)$ are functions of the cross-ratios $(\eta,\bar{\eta})$,
\begin{align}
    u = \eta \, \bar{\eta}, \quad v = (1-\eta)(1-\bar{\eta}).
\end{align}
In the previous section, the computation of the OTOC used the identity block approximation, which focused on the contribution of the unit operator and its Virasoro descendants, including the stress tensor. 
Further expanding in $1/c$, the ${\cal O}(1)$ contributions in \eqref{OTOC_line_beta} was due to the identity operator and the ${\cal O}(1/c)$ contribution arose from the stress tensor global block. In order to compute related contributions to the OTOC for a CFT on a spatial circle, it will be useful to draw intuition from a dual gravitational picture, in which global conformal blocks correspond to geodesic Witten diagrams. We will therefore briefly review the work of \cite{Hijano:2015zsa,daCunha:2016crm}, which relates the integral representation of conformal blocks in \citep{Ferrara:1971vh,Ferrara:1973vz,Ferrara:1974ny} to geodesic Witten diagrams.

In a 2d CFT, for operators with zero spin, $h=\bar{h}$, the global conformal block
takes the form  \cite{Ferrara:1971vh,Ferrara:1973vz,Ferrara:1974ny}
\begin{align}\label{integral_block}
    G_{h,h}(u,v) = \frac{1}{2\beta_h}u^h \int_0^1 & d\sigma \left(\sigma(1-\sigma)\right)^{h-1} (1-(1-v)\sigma)^{-h/2}\\
     & \times \; _2F_1\left( h, h, 2h, \frac{u \sigma (1-\sigma)}{1-(1-v)\sigma} \right)\nonumber.
\end{align}
Here the coefficient $\beta_h$ is the Euler beta function,
\begin{align}
    \beta_h = \frac{\Gamma(h)^2}{2\Gamma(2h)}.
\end{align}
This form explicit in cross-ratios is useful for computing the boundary $4$-point function. We have checked that using this form of the global conformal block, considering contributions from the identity and stress tensor global blocks and using the plane-to-cylinder transformation, we recover the expression of the OTOC in equation \eqref{OTOC_line_beta}. 

The bulk dual of a conformal block is the geodesic Witten diagram \cite{Hijano:2015zsa}, in which 
bulk vertices are integrated not over the whole bulk (as in ordinary Witten diagrams), but over geodesics connecting boundary points where CFT operators are located. In formulae, the geodesic Witten diagram is given by
\begin{align} \label{witten_diag}
& {\cal W}_{h,h}(z_i)  \equiv \\
& \int_{-\infty}^{\infty} d\lambda \int_{-\infty}^{\infty}  d\lambda' G_{b \partial }(y(\lambda),z_1) G_{b \partial }(y(\lambda),z_2) G_{bb} (y(\lambda), y(\lambda')) G_{b \partial }(y(\lambda'),z_3) G_{b \partial }(y(\lambda'),z_4),\nonumber
\end{align}
and the integral is over the geodesic parameters $\lambda, \lambda'$ of the geodesics $\gamma_{WW}, \gamma_{VV}$, respectively. Using the integral representation of the conformal block \eqref{integral_block}, \cite{Hijano:2015zsa} showed that the bulk dual of a conformal block is the geodesic Witten diagram,
\begin{align} \label{block_to_witten}
{\cal W}_{h,h}(z_i)  = \beta^2_h G_{h,h}(u,v).
\end{align}

Equation \eqref{witten_diag} can be further simplified by substituting the bulk-to-boundary propagators \cite{daCunha:2016crm}
\begin{align}
G_{b \partial}(y(\lambda),z_1)=  \frac{e^{ - \lambda \Delta_{W}}}{|z_{12}|^{\Delta_W}}, \quad G_{b \partial}(y(\lambda),z_2) = \frac{e^{  \lambda \Delta_{W}}}{|z_{12}|^{\Delta_W}},
\end{align}
and similarly for $G_{b \partial}(y(\lambda'),z_3)$ and $G_{b \partial}(y(\lambda'),z_4)$. Thus we obtain for the geodesic Witten diagram
\begin{align} \label{witten_bb}
 {\cal W}_{h,h}(z_i) = \frac{1}{|z_{12}|^{2\Delta_W} |z_{34}|^{2\Delta_V}} \int_{-\infty}^{\infty} d\lambda \int_{-\infty}^{\infty}  d\lambda' G_{bb} (y(\lambda), y(\lambda')).
\end{align}
In \cite{daCunha:2016crm}, the above form was used to interpret the geodesic Witten diagram, and thus the conformal block, as an integral of the bulk 2-point function of the bulk field $\Phi^{(0)}$ being exchanged,
\begin{align}
G_{bb} (y(\lambda), y(\lambda')) = \langle \Phi^{(0)} (y(\lambda))  \Phi^{(0)} (y(\lambda')) \rangle.
\end{align}
Here, the bulk field operator $\Phi^{(0)}$ is obtained from the boundary operator ${\cal O}$ (of conformal dimensions $h, \bar{h}$) by the HKLL prescription \citep{Hamilton:2006az}, and restricting to only the leading contribution in $1/N$ (i.e.\ ignoring multi-trace contributions \citep{Kabat:2011rz}). Thus the superscript~$^{(0)}$ indicates that it is a free bulk field. Substituting the above in \eqref{witten_bb} and \eqref{block_to_witten}, we obtain the boundary conformal block as an integral of the 2-point function of the bulk field dual to the corresponding boundary primary operator,
\begin{align} \label{block_bulk}
G_{h,h}(u,v) = \frac{1}{\beta^2_h |z_{12}|^{2\Delta_W} |z_{34}|^{2\Delta_V}} \int_{-\infty}^{\infty} d\lambda \int_{-\infty}^{\infty}  d\lambda' \langle \Phi^{(0)} (y(\lambda))  \Phi^{(0)} (y(\lambda')) \rangle.
\end{align}

An ingredient used implicitly in the above discussion is the operator product expansion of the boundary operators in terms of the bulk field. In \cite{Ferrara:1971vh, daCunha:2016crm} the contribution from primary operators ${\cal O}_{h,\bar{h}}$ to the OPE of a pair of boundary operators was expressed in terms of the bulk dual to ${\cal O}$,
\begin{align} \label{ope_bulk}
    W(z) W(0) \sim \frac{\beta_h^{-1}}{|x|^{4h_w}} \int_{-\infty}^\infty d\lambda \, \Phi^{(0)}(y(\lambda)).
\end{align}
While the above discussion is presented for the $4$-point function on a plane, obtaining the $4$-point function on the cylinder is straightforward by using the cylinder-to-plane conformal map \eqref{cyl_to_plane_map},
\begin{align}
    \langle  W(x_1,\bar{x}_1) & W(x_2,\bar{x}_2) V(x_3,\bar{x}_3) V(x_4,\bar{x}_4) \rangle_{\beta} \\
    & = \left| \frac{\partial z_1}{\partial x_1} \frac{\partial z_2}{\partial x_2} \right|^{2h_w}\left| \frac{\partial z_3}{\partial x_3} \frac{\partial z_4}{\partial x_4} \right|^{2h_v} \frac{1}{|z_{12}|^{4h_w}|z_{34}|^{4h_v}}\sum_{{\cal O}} \lambda_{WW {\cal O}} \lambda_{VV {\cal O}} G^{\beta}_{h,\bar{h}}(u,v).\nonumber
\end{align}
Here the superscript $\beta$ means that the arguments $(u,v)$ of the functions are written in terms of cylinder co-ordinates using the transformation \eqref{cyl_to_plane_map}. The cylinder-to-plane map can be extended to the bulk, relating Euclidean Poincar\'e AdS to Euclidean BTZ black branes \cite{Maldacena:1998bw, daCunha:2016crm}. This leads to a similar expression for $G^{\beta}_{h,\bar{h}}(u,v)$ as \eqref{block_bulk}, where the bulk field $\Phi^{(0)}$ is a function of co-ordinates in a boosted BTZ geometry instead of in AdS as was the case in \eqref{block_bulk}. To be explicit about this difference we write the bulk field $2$-point function with a subscript $\beta$,
\begin{align} \label{BTZ_block_bulk}
     G_{h,h}^\beta(u,v)  \propto \frac{4}{\beta^2_h} \int_{-\infty}^\infty d\lambda \int_{-\infty}^\infty d\lambda' \, \langle \Phi^{(0)}(y(\lambda)) \Phi^{(0)}(y(\lambda'))  \rangle_\beta.
\end{align}

%%%%%
\subsection{OTOC for 2d CFT on a spatial circle} \label{sec:CFT_circle}
\begin{figure}
    \centering
    \includegraphics[scale=1]{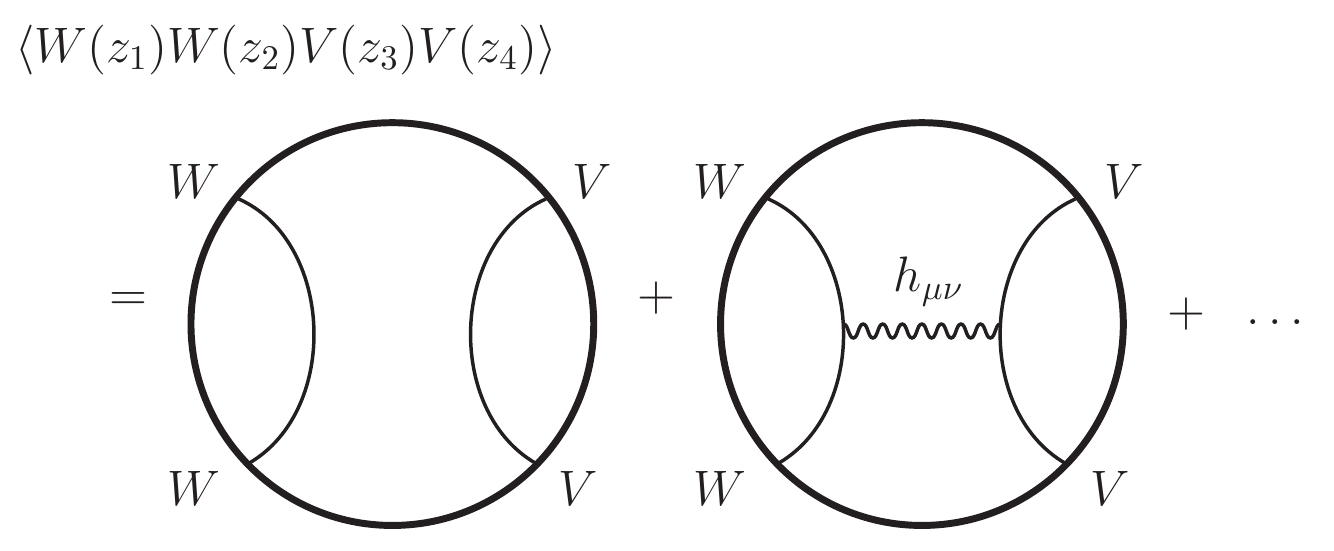}
    \caption{Disconnected and single-graviton-exchange geodesic Witten diagrams contributing to the OTOC.}
    \label{fig:witten_ads}
\end{figure}

Since we know that only the identity and stress tensor global blocks contribute to the OTOC at the leading orders in the $1/c$ expansion, in terms of the geodesic Witten diagram, the OTOC is a sum of a disconnected diagram and a graviton exchange diagram (see figure~\ref{fig:witten_ads}). In the latter a propagator of the graviton $h_{\mu \nu}$ connects the geodesics $\gamma_{VV}, \gamma_{WW}$ between the $VV$ and $WW$ boundary operators, respectively. Thus, the normalised boundary $4$-point function in the large $c$ limit can be written as the following function of the graviton $2$-point function in the bulk:
\begin{align}
   & \frac{\langle  W(x_1,\bar{x}_1) W(x_2,\bar{x}_2) V(x_3,\bar{x}_3) V(x_4,\bar{x}_4) \rangle}{\langle  W(x_1,\bar{x}_1) W(x_2,\bar{x}_2) \rangle \langle V(x_3,\bar{x}_3) V(x_4,\bar{x}_4) \rangle} \nonumber\\
    & \qquad \qquad = 1 + \lambda_{WW T} \lambda_{VV T} \int_{\gamma_{WW}} \int_{\gamma_{VV}} \langle h_{\mu \nu}(y(\lambda)) h_{\mu \nu}(y(\lambda')) \rangle.
\end{align}

However, our goal is to obtain the $4$-point function of boundary operators on the torus, corresponding to a CFT at finite temperature on a spatial circle. At least for heavy operators inserted close to each other, one expects that the OPE \eqref{ope_bulk} of the boundary operators in terms of the bulk field
can be used when evaluating $4$-point function on the torus. (Possible contributions due to non-minimal geodesics connecting the two operators would then be suppressed.)
Consider the case where the spatial periodicity is large compared to the time periodicity of the torus, and compared to the separation between the two $W$ operators and between the two $V$ operators. Then using \eqref{ope_bulk}, we obtain for the torus $4$-point function
\begin{align}\label{torus_bulk}
   & \frac{\langle  W(x_1,\bar{x}_1) W(x_2,\bar{x}_2) V(x_3,\bar{x}_3) V(x_4,\bar{x}_4) \rangle_{\beta,\ell}}{\langle  W(x_1,\bar{x}_1) W(x_2,\bar{x}_2) \rangle_{\beta,\ell} \langle V(x_3,\bar{x}_3) V(x_4,\bar{x}_4) \rangle_{\beta,\ell}} \nonumber \\
    & \qquad \qquad = 1 + \lambda_{WW T} \lambda_{VV T} \int_{\gamma_{WW}} \int_{\gamma_{VV}} \langle h_{\mu \nu}(y(\lambda)) h_{\mu \nu}(y(\lambda')) \rangle_{\beta,\ell}+\ldots 
\end{align}

In \citep{KeskiVakkuri:1998nw}, it was shown, using the method of images, that 2-point functions in a BTZ black hole (i.e.\ with compact spatial direction $\varphi \sim \varphi+ 2\pi$) can be written as a sum of 2-point functions in a BTZ black-brane geometry (i.e.\ the spatial direction is an infinite line), 
\begin{align}
    \langle {\cal O}(\varphi) \, {\cal O}(0) \rangle_{\beta,\ell} 
    = \sum_{n=-\infty}^\infty \langle {\cal O}(\varphi+2\pi n) \, {\cal O}(0) \rangle_{\beta}.
\end{align}
Here, the argument $\varphi$ on the LHS of the above equation is periodic, while it is not periodic in individual terms on the RHS. However the sum on the RHS has the effect of capturing the $\varphi$ periodicity of the LHS. The subscript 
on the RHS denotes that we are working with correlators at finite temperature. Using this equation in \eqref{torus_bulk}, we can write the graviton 2-point function on the torus as
\begin{align}\label{torus_graviton_propagator}
    \langle h_{\mu \nu}(y(\lambda)) h_{\mu \nu}(y(\lambda')) \rangle_{\beta,\ell} = \sum_{n=-\infty}^\infty \langle h_{\mu \nu, n}(y(\lambda)) h_{\mu \nu}(y(\lambda')) \rangle_{\beta},
\end{align}
where $h_{\mu \nu, n}$ denotes that the $\varphi$ argument is shifted by $2\pi n$.

Substituting equations \eqref{torus_graviton_propagator} and \eqref{BTZ_block_bulk} in \eqref{torus_bulk}, we obtain the following expression for the normalised torus $4$-point function,
\begin{align} \label{torus_sum_images}
    &{\cal C}_{\beta,\ell}(\sigma,t) \equiv \frac{\langle  W(x_1,\bar{x}_1) W(x_2,\bar{x}_2) V(x_3,\bar{x}_3) V(x_4,\bar{x}_4) \rangle_{\beta,\ell}}{\langle  W(x_1,\bar{x}_1) W(x_2,\bar{x}_2) \rangle_{\beta,\ell} \langle V(x_3,\bar{x}_3) V(x_4,\bar{x}_4) \rangle_{\beta,\ell}} \nonumber\\
    & \qquad \qquad = 1 + \lambda_{WW T} \lambda_{VV T} \sum_{n=-\infty}^\infty G^\beta_{2,2}(u_n,v_n) +\dots.
\end{align}
The second term on the RHS is a sum over stress tensor global blocks on the thermal cylinder. The subscript $n$ in the cross-ratios $u_n,v_n$ means that the spatial separation (in $\varphi$) between $W$ and $V$ operators is shifted by $2\pi n$.\footnote{On the torus, $n$ labels the number of times the graviton winds the spatial circle. Torus conformal blocks have definite winding numbers \cite{Kraus:2017ezw}, so an infinite number of torus blocks contribute to the sum.}

In order to compute the OTOC from this $4$-point function, we use the locations of the operators on the plane,
\begin{align} \label{OTOC_loc_n}
z_{1} & = e^{\frac{2\pi}{\beta_L} \left(i\epsilon_1 \right)} \qquad \qquad \qquad \quad \;\; \bar{z}_{1} = e^{\frac{2\pi}{\beta_R}\left(-i\epsilon_1\right)} \nonumber\\
z_{2} & =e^{\frac{2\pi}{\beta_L} \left(i\epsilon_2 \right)} \qquad \qquad \qquad \quad \;\; \bar{z}_{2} = e^{\frac{2\pi}{\beta_R}\left(-i\epsilon_2\right)}\nonumber\\
z_{3,n} & = e^{\frac{2\pi}{\beta_L}\left(\sigma-2\pi n \ell-t+i\epsilon_3 \right)} \qquad \; \; \; \, \bar{z}_{3,n} = e^{\frac{2\pi}{\beta_R}\left(\sigma-2\pi n \ell+t-i\epsilon_3\right)} \nonumber\\
z_{4,n} & = e^{\frac{2\pi}{\beta_L}\left(\sigma-2\pi n \ell-t+i\epsilon_4 \right)} \qquad \; \; \; \, \bar{z}_{4,n} = e^{\frac{2\pi}{\beta_R}\left(\sigma-2\pi n \ell+t-i\epsilon_4\right)},
\end{align}
where the thermal cylinder-to-plane map has been used.
The cross-ratios are
\begin{align}
u_{n} = \eta_{n} \, \bar{\eta}_{n}, \qquad v_{n} =\left( 1-\eta_{n}\right) \left( 1-\bar{\eta}_{n}\right),
\end{align}
where
\begin{align} 
\eta_{n} = \frac{-\sin \left( \frac{\pi}{\beta_L}\epsilon_{12} \right) \sin \left( \frac{\pi}{\beta_L}\epsilon_{34} \right)}{\sinh \left(  \frac{\pi}{\beta_L}\left(t- \sigma + 2\pi \ell n +i\epsilon_{13} \right)\right) \sinh \left(  \frac{\pi}{\beta_L}\left(t- \sigma + 2\pi \ell n +i\epsilon_{24} \right)\right)}\\
\bar{\eta}_{n} = \frac{-\sin \left( \frac{\pi}{\beta_R}\epsilon_{12} \right) \sin \left( \frac{\pi}{\beta_R}\epsilon_{34} \right)}{\sinh \left(  \frac{\pi}{\beta_R}\left(t + \sigma - 2\pi \ell n +i\epsilon_{13} \right)\right) \sinh \left(  \frac{\pi}{\beta_R}\left(t + \sigma - 2\pi \ell n +i\epsilon_{24} \right)\right)}.
\end{align}
The cross-ratios in the large time limit $t \gg |\sigma - 2\pi \ell n|, \beta_{L/R}$ are
\begin{align} \label{eta_circle_larget}
\eta _{n} & = -\tilde{\epsilon}^{*L}_{12}\tilde{\epsilon}^{L}_{34} e^{- \frac{2\pi}{\beta_L}\left(t- \sigma + 2\pi \ell n \right)} \nonumber\\
\bar{\eta} _{n} & = -\tilde{\epsilon}^{*R}_{12}\tilde{\epsilon}^{R}_{34} e^{- \frac{2\pi}{\beta_R}\left(t + \sigma - 2\pi \ell n \right)}.
\end{align}

The conformal blocks appearing in \eqref{block_expansion} are known in even dimensions in terms of hypergeometric functions \citep{Dolan:2011dv}, and for $d=2$ they take the form
\begin{align}
    G_{h,\bar{h}}(u,v) = \eta^h \bar{\eta}^{\bar{h}} \, _2F_1(h,h,2h,\eta) \, _2F_1(\bar{h},\bar{h},2\bar{h},\bar{\eta}). \nonumber
\end{align}
Since the hypergeometric function has known monodromies at $\eta=1$, the stress tensor global block $G^\beta_{2,2}(u,v)$
has a branch cut from $\eta, \bar{\eta} = [1,\infty)$. As a consequence, the block can get contributions from multiple channels, as discussed in section~\ref{sec:line}. To obtain the correct correlator, we choose the channel which gives the dominant contribution to the OTOC.
Thus for each term in the sum appearing in \eqref{torus_sum_images},
\begin{align} \label{channel_choice}
    \tilde{G}^\beta_{2,2}(u_n,v_n) \equiv \lambda_{WWT} \lambda_{VVT} \, G^\beta_{2,2}(u_n,v_n) \approx  -   \frac{48\pi i  h_w h_v}{c \eta_{n}}  \quad \mathrm{or}\quad  -\frac{48\pi i  h_w h_v}{c \bar{\eta}_{n}} \,.
\end{align}
Using the late time limit of the cross-ratios in equation \eqref{eta_circle_larget}, and comparing $\tilde{G}^\beta_{2,2}(u_n,v_n)$ in the two channels, the channel in which this quantity is smaller contributes to the OTOC:%
\footnote{This might appear at odds with our prescription in section~\ref{sec:line}, where we maximised over channels. Note, however, that we were maximising Virasoro blocks like~\eqref{otoc_line_hhll}, which corresponds to minimising the exponential terms in~\eqref{OTOC_line_beta}. For CFT on a spatial circle we do not have expressions for Virasoro blocks, and we assume that we should still minimise the stress tensor global blocks.}
\begin{align}
    \tilde{G}^\beta_{2,2}(u_n,v_n) \approx
    \begin{cases}
     - \frac{48\pi i h_w h_v}{c \,\tilde{\epsilon}^{*L}_{12}\tilde{\epsilon}^{L}_{34}} e^{ \frac{2\pi}{\beta_L}\left(t- \sigma + 2\pi \ell  n \right)}, &  n < n^*\\
      - \frac{48\pi i h_w h_v}{c \,\tilde{\epsilon}^{*R}_{12}\tilde{\epsilon}^{R}_{34}} e^{ \frac{2\pi}{\beta_R}\left(t + \sigma - 2\pi \ell  n \right)}, &  n \geq n^*,
    \end{cases}
\end{align}
where
\begin{align} \label{nstar}
    n^* \equiv \frac{\sigma^*-\Omega \, t}{2\pi \ell},
\end{align}
and $\sigma^*$ is defined in equation \eqref{sigma_star}.
Substituting the above equation in \eqref{torus_sum_images} we find for the OTOC on the torus
\begin{align} \label{OTOC_sum_exp}
 {\cal C}_{\beta,\ell}(\sigma,t) = 1-    \frac{48\pi i h_w h_v}{c}    e^{\frac{2\pi \, t}{\beta}} & \bigg( \sin^{-1} \left( \frac{\pi \epsilon_{12} }{\beta_L} \right) \sin^{-1} \left( \frac{\pi \epsilon_{34}}{\beta_L} \right) e^{-\frac{2\pi}{\beta}\frac{\ell \phi}{(1-\Omega)}}\sum_{n=-\floor{ \frac{t-\sigma}{2\pi \ell}}}^{\floor{n^*}} e^{\frac{4\pi^2 \ell n}{\beta (1-\Omega)}}  \nonumber\\
&   + \sin^{-1} \left( \frac{\pi \epsilon_{12} }{\beta_R} \right) \sin^{-1} \left( \frac{\pi \epsilon_{34}}{\beta_R} \right) e^{\frac{2\pi}{\beta}\frac{\ell \phi}{(1+\Omega)}}\sum_{n=\lceil n^*\rceil}^{\floor{\frac{t+\sigma}{2\pi \ell}}} e^{-\frac{4\pi^2 \ell  n}{\beta (1+\Omega)}}  \bigg).
\end{align}
The lower limit in the first sum and the upper limit in the second sum are imposed because the continuation across the branch cut in Lorentzian time
only occurs when $t>|\sigma - 2\pi \ell n|$. After performing the sums, we get
\begin{align} \label{OTOC_circle_beta}
 & {\cal C}_{\beta,\ell}(\sigma,t) =\\
 &    1-   \frac{48\pi i h_w h_v}{c }  e^{\frac{2\pi \, t}{\beta}}  \left( \sin^{-1} \left( \frac{\pi \epsilon_{12} }{\beta_L} \right) \sin^{-1} \left( \frac{\pi \epsilon_{34}}{\beta_L} \right) \left( \frac{1+\Omega}{1-\Omega}\right)^{1+\Omega}\frac{  \exp \left( -\frac{4\pi^2}{\beta}\frac{\ell (n^* \, \text{mod}\, 1)}{(1-\Omega)} \right)}{1-\exp \left(- \frac{4\pi^2 \ell}{\beta (1-\Omega)} \right)} \right. \nonumber\\
& \qquad \qquad \qquad \quad \quad \left. +\sin^{-1} \left( \frac{\pi \epsilon_{12} }{\beta_R} \right) \sin^{-1} \left( \frac{\pi \epsilon_{34}}{\beta_R} \right) \left( \frac{1-\Omega}{1+\Omega}\right)^{1-\Omega}\frac{ \exp \left(\frac{4\pi^2}{\beta}\frac{\ell (n^* \, \text{mod}\, 1)}{(1+\Omega)} \right)}{\exp \left( \frac{4\pi^2 \ell}{\beta (1+\Omega)} \right)-1} \right).\nonumber
\end{align}
For small $\epsilon_i$, the OTOC becomes
\begin{align}
    & {\cal C}_{\beta,\ell}(\sigma,t)= 1-\\
    & \frac{48 i h_w h_v \beta^2(1+\Omega)^{1+\Omega} (1-\Omega)^{1-\Omega}}{\pi \, c \, \epsilon_{12} \epsilon_{34} }   e^{\frac{2\pi \, t}{\beta}}  \left( \frac{  \exp \left( -\frac{4\pi^2}{\beta}\frac{\ell (n^* \, \text{mod}\, 1)}{(1-\Omega)} \right)}{1-\exp \left(- \frac{4\pi^2 \ell}{\beta (1-\Omega)} \right)} +\frac{ \exp \left(\frac{4\pi^2}{\beta}\frac{\ell (n^* \, \text{mod}\, 1)}{(1+\Omega)} \right)}{\exp \left( \frac{4\pi^2 \ell}{\beta (1+\Omega)} \right)-1} \right).\nonumber
\end{align}
Comparing this to the result obtained in  \citep{Mezei:2019dfv}, we find, as in section~\ref{sec:line}, a shift in time as well as the prefactors which were not determined in \citep{Mezei:2019dfv}.
As was pointed out in \citep{Mezei:2019dfv}, the OTOC decays in time as an exponential with average Lyapunov exponent $\bar{\lambda}_L = \frac{2\pi}{\beta}$. However, it has periodic modulations about this average behaviour captured by $(n^* \, \text{mod} \, 1)$. Also note that in the decompactification limit, $\ell \to \infty$ keeping $\sigma =\ell \varphi$ constant, using $\phi = \frac{\sigma-\Omega t}{\ell} \to 0$, we recover the OTOC as obtained in equation \eqref{OTOC_line_beta}.
 
 %%%%%%%%
 \subsection{Comment on ensembles}\label{subsec:emsembles}
As mentioned, for instance, in \cite{Mezei:2019dfv}, where the OTOC of interest was computed from a bulk point of view, the BTZ black hole only dominates the canonical ensemble for sufficiently high temperature, 
\begin{align}\label{temperature_bound}
    \frac{\beta}{\ell} < \frac{2\pi}{\sqrt{1-\Omega}}.
\end{align}
For lower temperatures, the thermal gas in AdS dominates \citep{Hawking:1982dh,Hartman:2014oaa}. This raises a few questions:
\begin{enumerate}
    \item For lower temperatures, do the computations of OTOCs in BTZ performed in \cite{Mezei:2019dfv} have a CFT counterpart? This question applies even more strongly to the bulk computations of \cite{Craps:2020ahu}, which focused on extremal BTZ.
    \item Our CFT computation of the OTOC would seem to agree with the BTZ result of \cite{Mezei:2019dfv} even at low temperatures -- how can this be the case if the BTZ black hole does not dominate the ensemble?
\end{enumerate}
Let us consider these questions in the context of the D1-D5 CFT, which is often used for holographic studies of black holes in string theory \cite{Strominger:1996sh}. If one really takes a thermal average over all states in the theory, then it is indeed true that black hole states play no significant role at low temperatures. For instance, at zero temperature the free energy coincides with the energy, and is therefore minimized by the ground state of the theory, which is the NS vacuum, whereas the R ground states (which include the lightest black hole states) have higher energy. But this suggests a modification of the CFT computation that does capture black hole physics, namely one can restrict the thermal average to R sector states. For low temperatures that violate \eqref{temperature_bound}, this is what we will implicitly assume.\footnote{We thank I.~Bena, M.~De Clerck, F.~Denef, K.~Nguyen, R.~Russo and N.~Warner for discussions related to this point.}

But what if we did not assume a restriction to R sector states? Don't our CFT computations generally reproduce BTZ results? The resolution is that for CFT on a torus there are different channels in which one can perform an OPE expansion, and for most channels a truncation of a correlation function to the lowest few terms will not provide a good approximation to the exact result. By starting from geodesic Witten diagrams for a BTZ black brane and using the method of images for the bulk-to-bulk propagator, we implicitly chose the channel whose truncation works well in the high-temperature phase. If one considered low temperatures and included the NS sector states, the ``good'' channel would correspond to summing over states propagating along the other cycle of the torus, which is a computation we did not perform.

In the next section, we consider zero-temperature limits of our results. We will work in the ``black hole channel'' and implicitly assume that we are restricting the thermal trace to R sector states. 

%%%%%%%%%%%%%%%%%%%%%%%%%%%%%%%%%%%%%%%%%%%%%%%%%%%%%%%%%%%%%%%%%%%%%%%%%%%%%%%%%%%%%%%%%%%%%%%%%%%%%%%%%%%%%
\section{Extremal limits}\label{sec:extremal}
In recent work \citep{Craps:2020ahu}, the bulk computations of \citep{Mezei:2019dfv} have been extended to extremal, maximally rotating BTZ black holes, motivated in part by the question to what extent the OTOC can distinguish between those black holes and their horizonless microstate geometries.
In order to compare the results of \citep{Craps:2020ahu} with CFT computations, in this section we extend the OTOC computations of sections~\ref{sec:line} and~\ref{sec:CFT_circle} to the extremal, maximally rotating limit, $\Omega \to 1$, $\beta \to \infty$,
\begin{align}
\beta_L = \text{finite}, \qquad \beta_R \to \infty.
\end{align}

\subsection{CFT on the line}
To obtain the OTOC in the extremal limit for the computation in section~\ref{sec:line} we cannot simply take $\beta_R \to \infty$ in \eqref{OTOC_line_beta}. This is because this limit competes with the $t \gg |\sigma|, \beta$ limit which was used to arrive at the OTOC at finite temperature. In order to obtain the OTOC in the extremal limit, we consider $\beta_R \to \infty$ in the cross-ratio before substituting the cross-ratio in \eqref{eqn:OTOC_candidate_channels}. 
Following the reasoning of section~\ref{sec:line}, we find that for $t >|\sigma|$,
\begin{align} \label{eq:OTOC_extremal_decompact}
    {\cal C}(\sigma,t) & \approx 1- \frac{48 \pi i h_w h_v}{c} \frac{
\sinh^2 \left( \frac{\pi}{\beta_L} \left(t- \sigma\right) \right)}{ \sin \left( \frac{\pi \epsilon_{12}}{\beta_L}\right)  \sin \left( \frac{\pi \epsilon_{34}}{\beta_L} \right) }, \quad  \frac{\beta_L}{\pi} \sinh \left( \frac{\pi}{\beta_L} \left(t- \sigma\right) \right) < t+\sigma \nonumber\\
& \approx 1- \frac{48 \pi i h_w h_v}{c \, \epsilon_{12} \epsilon_{34}} \left(t+\sigma\right)^2 ,\qquad \qquad \quad  \frac{\beta_L}{\pi} \sinh \left( \frac{\pi}{\beta_L} \left(t- \sigma\right) \right) \geq t+\sigma.
\end{align}

\subsection{CFT on the circle}
In the maximally rotating limit, for a given $n$, the two channels described in equations \eqref{channel_choice} give the following contributions,
\begin{align}
    \tilde{G}^\beta_{2,2}(u_n,v_n) \approx  - \frac{48 \pi i h_w h_v}{c} \min \left(
   \frac{\sinh^2 \left( \frac{\pi}{\beta_L} \left(t- \sigma -2\pi \ell \tilde{n} \right) \right)}{ \sin\left( \frac{\pi \epsilon_{12}}{\beta_L}\right) \sin \left( \frac{\pi \epsilon_{34}}{\beta_L} \right) },
   \frac{ \left(t+\sigma + 2\pi \ell \tilde{n} \right)^2}{\epsilon_{12}\epsilon_{34}}\right),
\end{align}
where in the above equation we have defined $\tilde{n} \equiv -n$.
The cross-over takes place at 
\begin{align}
  \frac{\beta_L^2}{\pi^2} \sinh^2 \left( \frac{\pi}{\beta_L} \left(t- \sigma -2\pi \ell \tilde{n}_* \right) \right) = \left(t+\sigma + 2\pi \ell \tilde{n}_* \right)^2,
\end{align}
and the value of $\tilde{n}$ at the cross-over point is denoted as $\tilde{n}_*$. Solving the above equation for $\tilde{n}_*$,
\begin{align}
    \tilde{n}_* \approx \frac{1}{2\pi \ell} \left( t - \sigma - \frac{\beta_L}{\pi}\log(\frac{4\pi t}{\beta_L}) \right).
\end{align}
Substituting this in \eqref{torus_sum_images}, we obtain for the OTOC on a spatial circle in the extremal limit
\begin{align}
    {\cal C}_{\beta,\ell}(\sigma,t) = 1 - \frac{48\pi i h_w h_v}{c \, \epsilon_{12} \epsilon_{34}} \left( \frac{\beta_L^2}{\pi^2}   {\sum_{\tilde{n}= \lceil \tilde{n}_* \rceil }^{\lfloor (t-\sigma)/(2\pi \ell) \rfloor}}  \sinh^2 \left( \frac{\pi}{\beta_L} \left(t- \sigma -2\pi \ell \tilde{n} \right) \right) \right. \nonumber\\
     \left. + \sum_{\tilde{n}=-\lfloor (t+\sigma)/(2\pi \ell) \rfloor}^{ \lfloor \tilde{n}_* \rfloor}  \left(t+\sigma + 2\pi \ell \tilde{n}\right)^2 \right).
\end{align}
Here the upper and lower bounds on the first and second sum, respectively, take into account that these contributions are only present for $t> |\sigma + 2\pi \ell \tilde{n}|$. The sums appearing in the above equations are like those encountered in equation (3.23) of \citep{Craps:2020ahu} with their \lq$ n$\rq \, the same as our $\tilde{n}$ and their \lq$r_+$\rq \, replaced by our $\frac{\pi}{\beta_L}$. Evaluating the sums, we obtain the OTOC
\begin{align}
    {\cal C}_{\beta,\ell}(\sigma,t) \approx & 1 - \frac{24 \pi i h_w h_v }{ c\,  \epsilon_{12} \epsilon_{34}}  \left[\frac{2 t^3}{3\pi \ell}-\frac{ t^2 \beta_L \log \left(\frac{4 \pi  t}{\beta_L}\right)}{\pi^2 \ell}  + \left(1-\frac{  \left[\frac{\pi  (t-\sigma)-\beta_L \log \left(\frac{4 \pi  t}{\beta_L}\right)}{\ell \pi } \bmod 2 \pi \right]}{ \pi}\right.\right. \nonumber\\
    &\left. \left. + \frac{2 \exp \left(\frac{2 \pi  \ell }{\beta_L}\left[\frac{\pi  (t-\sigma)-\beta_L \log \left(\frac{4 \pi  t}{\beta_L}\right)}{\ell \pi } \bmod 2 \pi \right]\right)}{e^{\frac{4 \pi ^2 \ell}{\beta_L}}-1} \right)t^2 + {\mathrm{O}\left(\frac{t}{\ell} \right)}\right].
\end{align}
Thus the OTOC grows as $t^3$ on an average with a saw-tooth pattern superimposed where there are brief periods of $t^2$ growth and brief periods of exponential growth dictated by the non-zero left temperature $1/\beta_L$. This behaviour of the OTOC in the extremal limit agrees with that obtained from the bulk in \citep{Craps:2020ahu}. 

\paragraph{Acknowledgements}
We would like to thank I.~Bena, M.~De Clerck, F.~Denef, P.~Hacker, M.~Hughes, K.~Nguyen, R.~Russo and N.~Warner for discussions and/or collaboration on related work. This research has been supported by FWO-Vlaanderen project G006918N and by Vrije Universiteit Brussel through the Strategic Research Program High-Energy Physics. CR was supported by FWO-Vlaanderen postdoctoral fellowship 12ZQ320N.

%%%%%%%%%%%%%%%%%%%%%%%%%%%%%%%%%%%%%%%%%%%%%%%%%%%%%%%%%%%%%%%%%%%%%%%%%%%%%%%%%%%%%%%%%%%%%%%%%%%%
\appendix

%%%%%%%%%%%%%%%%%%%%%%%%%%%%%%%%%%%%%%%%%%%%%%%%%%%%%%%%%%%%%%%%%%%%
\section{OTOC in vacuum of 2d CFT on spatial circle}\label{app:vacuum}
In \citep{Roberts:2014ifa}, the OTOC was computed in the vacuum of a 2d CFT on a line. In section~\ref{sec:extremal} of the present paper, we computed the OTOC in an ensemble with zero right-moving and nonzero left-moving temperature, both on a line and on a circle. For completeness, in this appendix we compute the OTOC in the vacuum of a 2d CFT on a spatial circle.

In the case when the spatial direction of the CFT is non-compact, the cross-ratios at zero temperature are
\begin{align}
\eta \approx - \frac{\epsilon_{12}\epsilon_{34}}{(t-\sigma)^2}, \qquad \bar{\eta} \approx - \frac{\epsilon_{12}\epsilon_{34}}{(t+\sigma)^2}.
\end{align}
Substituting the cross-ratios in equations \eqref{eqn:OTOC_candidate_channels}, the following OTOC is obtained \citep{Roberts:2014ifa}:
\begin{align}
{\cal C}(\sigma,t)  \approx 1- \frac{48\pi i h_w h_v}{ c \, \epsilon_{12}\epsilon_{34}} 
\begin{cases}
({t-\sigma})^2, & \sigma >0,\\
({t+\sigma})^2, & \sigma\leq 0.
\end{cases}
\end{align}

In the CFT on a spatial circle discussed in section \ref{sec:CFT_circle}, in the zero temperature limit, $\beta \to \infty$, the cross-ratios are
\begin{align}
\eta_n \approx - \frac{(\epsilon_1-\epsilon_2)(\epsilon_3-\epsilon_4) }{\left(t-\sigma+2\pi n\ell \right)^2}, \qquad \bar{\eta}_n \approx - \frac{(\epsilon_1-\epsilon_2)(\epsilon_3-\epsilon_4)}{\left( t+\sigma-2\pi n\ell \right)^2}.
\end{align}
To obtain the OTOC, we substitute the above cross-ratios in the equivalent of equation \eqref{torus_sum_images} for vacuum and impose the constraint $t >|\sigma-2\pi n\ell|$ on the summations:   
\begin{align} \label{OTOC_sum_pol}
{\cal C}_{\rm vac}^{\rm circle}(\sigma,t)  & = 1+  \frac{48\pi i h_w h_v}{c }  \left( \sum_{n=-\floor{ \frac{t-\sigma}{2\pi \ell}}}^{\floor{\frac{\sigma}{2\pi \ell}}}  \eta_{n}^{-1} +\sum_{n=\lceil\frac{\sigma}{2\pi \ell}\rceil}^{\floor{\frac{t+\sigma}{2\pi \ell}}} \bar{\eta}_{n}^{-1} \right).
\end{align}
Thus, the OTOC on the torus at zero temperature 
becomes
\begin{align}
& {\cal C}_{\rm vac}^{\rm circle}(\sigma,t)\approx 1-\\ &  \frac{16\pi i h_w h_v \ell^2}{c \, \epsilon_{12}\epsilon_{34}} \bigg[ \frac{t^3}{\ell^3} + \frac{2\pi^2 t}{\ell} \left( 1 + 6 \left( -1 + \left( \frac{\sigma}{2\pi \ell} \bmod 1 \right)\right)\left( \frac{\sigma}{2\pi \ell} \bmod 1 \right) \right) +\mathrm{O} \left(\left( \frac{t}{\ell}\right)^0\right)  \bigg] . \nonumber
\end{align}

\bibliographystyle{JHEP}
\bibliography{references_OTOC.bib}

\end{document}